# Does time to retreatment matter? An NTCP model to predict radionecrosis after repeat SRS for recurrent brain metastases incorporating time-dependent discounted dose.

Manju Sharma[1] PhD, Issam El Naqa[2] PhD, Penny K Sneed[1] MD

[1]Department of Radiation Oncology, University of California, San Francisco, CA, USA,
[2]Department of Machine Learning, Moffitt Cancer Center, Tampa, FL, USA

**Purpose:** To develop and compare normal tissue complication probability (NTCP) models for recurrent brain metastases (BMs) treated with repeat single-fraction stereotactic radiosurgery (SRS), considering time-dependent discounted prior dose.

**Methods:** We developed three NTCP models of BMs treated with GammaKnife-based SRS. The maximum dose to 0.2cc ($D_{0.2cc}$) of each lesion-specific brain and one-year radionecrosis was fitted using a logistic model with equivalent-dose conversions in 2 Gy (EQD2). The M0 and M1-retreat modeled radionecrosis risk following SRS to 1029 non-recurrent lesions (patients=262) and 2nd SRS to 149 recurrent lesions (patients=87). The M1-combo model accounted for 2nd SRS and time-dependent discounted 1st SRS dose for recurrent lesions estimated by a modified Gompertzian function.

**Results:** All three models fitted the data well (Chi-2 = 0.039-0.089 and p = 0.999-1.000). The fitted $EQD2_{50}$ was ~103 Gy for M0, ~88 Gy for M1-retreat, and ~165 Gy for M1-combo. The fitted $\gamma_{50}$ exhibited a progressively flatter dose-response curve across the three models, with values of 1.2 Gy for M0, 0.6 Gy for M1-retreat, and 0.4 Gy for M1-combo. For the brain $D_{0.2cc}$ of 29Gy and 19Gy, the steepest to shallowest dose-response or largest change in NTCP, i.e., $NTCP_{29Gy}$ - $NTCP_{19Gy}$ was observed in M1-retreat (0.16), M0 (0.14) and M1-combo (0.06).

**Conclusions:** The model-fitted parameters predict that recurrent BMs have a lower threshold dose tolerance and a more gradual dose response for the 2nd SRS than non-recurrent BMs. This gradual dose-response becomes even more apparent when considering the time-dependent discounted 1st SRS as a cumulative 2nd SRS. Tailoring SRS retreatment protocols based on NTCP modeling can potentially enhance therapeutic efficacy.

# Introduction:

Stereotactic radiosurgery (SRS) is a well-established treatment option for patients with limited brain metastases (BMs) [1, 2]. However, local recurrence remains a risk, and while the use of a second course of SRS for recurrent BMs has increased over the last decade, its efficacy and tolerability are still debated. The competing risk factors include the need to escalate the dose to effectively target radio-resistant clones [3] and the substantial risk of adverse events, such as radiation-induced brain adverse radiation effect, commonly referred to as "radionecrosis."

An optimal retreatment prescription is contingent upon factors such as lesion volume, histology, number of recurrent lesions, location, prior radiation dose, and the anticipated dose-response dynamics for adverse effects [4]. Additionally, the time to recurrence can affect these factors, influencing retreatment doses based on the residual impact from initial treatment. The use of time-dependent dose discounts or forgiveness factors in recurrence settings is controversial, with clinicians often relying on institutional practices rather than well-defined clinical-based evidence. There is considerable variability in SRS retreatment doses and fractionation [5]. Some studies used a reduced dose [6, 7], others maintained the same dose [8], and some increased the dose [9]. Similar doses were prescribed at our institution for retreatment and the initial treatment.

Experimental and clinical data reveal significant variations in tissue capacity for long-term radiation injury recovery. While some tissues, such as the heart, bladder, and kidney, show no long-term recovery from radiation injury, others, including the brain, skin, mucosa, lung, and spinal cord, exhibit a recovery process over time [10, 11]. The potential for reirradiation to achieve maximal therapeutic benefit depends on understanding the radioresistance of lesions and the recovery process from radiation damage of normal tissues. Incorporating elapsed time and tissue-specific recovery patterns in normal tissue complication probability (NTCP) is crucial to accurately predict the risk of adverse effects following repeat radiation treatment. This approach allows for improved decision-making regarding retreatment alternatives. Without considering these factors, reirradiation may result in sub-optimal outcomes and significant toxicities, such as symptomatic radionecrosis, which may severely impact the patient's quality of life. [11-14]

We aimed to understand the risk of brain radionecrosis after repeat single-fraction SRS for recurrent lesions. We hypothesized that the NTCP would differ among BMs undergoing re-irradiation for recurrence and those without re-irradiation. Necrosis risk has been shown to correlate with volume and dose metrics [15-17]. We opted for dose metrics for their practicality when time-dependent discounted prior doses are applied. Various studies have found that the maximum dose received by small brain volumes predicts radionecrosis [15, 16, 18]. Using the logistic model suggested by the HyTEC project [19], we modeled NTCP for radionecrosis after single-fraction treatments. After testing various dose metrics, we selected $D_{0.2cc}$ for its best chi-squared fit for our models. Our NTCP prediction model estimated brain radionecrosis using the $D_{0.2cc}$ dose to normal brain tissue adjacent to the lesion, with three distinct models: no-recurrence, recurrence with repeat SRS dose only, and recurrence with cumulative dose accounting for initial time-dependent discounted SRS dose and repeat SRS dose. We employed a modified Gompertzian function with a threshold to model tissue response dynamics post-irradiation, explicitly focusing on time-dependent dose discounts. This approach accounts for the time-dependent recovery of normal tissue, aiming to quantitatively understand the dose-response relationship in cases of recurrence while considering prior dose effects. To our knowledge, this is the first attempt to quantitatively understand the time effect in the normal tissue dose response by utilizing the largest reported single-institution patient cohort with and without recurrence.

## Materials and Methods:

The first step of our study involved constructing an NTCP prediction model for non-recurrent BM lesions using the HyTEC methodology [19], denoted as Model M0. Subsequently, we developed two distinct NTCP models for the recurrent dataset. Model M1-retreat mirrored M0 but accounted solely for the dose from the repeat SRS. Model M1-combo incorporated the time-discounted initial SRS dose with the second SRS dose and additional WBRT doses where applicable.

**Patient Data:**

Our group previously studied adverse radiation effects (ARE) in patients who underwent Gamma Knife SRS for BMs treated from 1998 to 2009 [13] and in recurrent BMs treated with repeat SRS from 1998 to 2019 [12], following Institutional Review Board approval. Standard follow-up included a brain MRI every three months. For each patient, dates of the last brain MRI, death, or last follow-up were recorded. Progression dates and AREs (asymptomatic and symptomatic) were recorded for each lesion. Exclusion criteria for the original dataset are detailed in references [3-4]. In summary, patients were excluded if they had less than three months of follow-up imaging after SRS or gaps exceeding eight months in imaging in the first year after SRS, which could result in undetected radiographic ARE. This study is limited to symptomatic ARE, called radionecrosis in this study for simplicity. It specifically includes lesions treated with single-fraction SRS but excludes those with incomplete initial or retreatment radiation plans, less than 12 months of follow-up imaging, or incomplete follow-up data. Table 1 lists the characteristics of the two cohorts: the non-recurrent cohort included 262 patients with 1029 metastatic brain lesions, while the recurrent cohort had 87 patients with 149 metastatic brain lesions.

**Treatment methodology:**

Patients treated before 2001 received radio-surgical treatment using the Model B Leksell Gamma Knife, while those from 2001 to 2007 were treated with the Model C or 4C and then transitioned to the Perfexion Gamma Knife starting in November 2007. On the day of SRS, a Leksell Coordinate Frame G was affixed by the neurosurgeon under local anesthesia, followed by the contrast-enhanced T1-weighted MRI scans, target delineation, and treatment planning using

Leksell GammaPlan software (Elekta AB). The planning target volume (PTV) was the same as the gross tumor volume (GTV). The prescribed dose, determined based on treatment volume, ranged broadly from ~15-20 Gy, with smaller lesions receiving lower doses. Follow-up brain MRI every three months was recommended, and patients who underwent imaging elsewhere were asked to send the images to the Gamma Knife Coordinator for review at our weekly SRS conference for consistency and to assess for ARE or local treatment failure [12, 13].

## Scoring of Radionecrosis:

Serial imaging had been re-reviewed by one of the authors (XXX) in preparation for previous publications [12, 13] to score local failure and ARE ("radionecrosis") by lesion, taking into account the entire period of imaging follow-up as well as pathology results in case of surgical intervention. An event was scored in case of imaging showing at least a 2 mm increase in BM diameter or at least a 25% increase in estimated ellipsoid volume for irregular lesions. Progressive enlargement was interpreted as local failure, whereas stability or shrinkage over time was interpreted as radionecrosis [12, 13].

## Modeling Data:

The input modeling parameters included brain dose metrics ($D_{0.1cc}$, $D_{0.2cc}$, and $D_{0.3cc}$) and the binary clinical radionecrosis data. Radiosurgical treatment was delivered using different models of the Leksell Gamma Knife over the years. As such, all Gamma Knife plans were recalculated using the GammaPlan TMR10 dose calculation algorithm with a 1mm dose resolution grid. The DICOM-RT MRI images, treatment plans, and RT structure files were exported to MIM software V7.4. A lesion-specific brain region of interest (ROI) was created in MIM using Boolean operations. This involved expanding the lesion by a few millimeters, taking the union of the expanded structure with the regional 10 Gy isodose line structure, and then subtracting the lesion volume. Lesions with overlapping 10 Gy isodose lines were excluded due to the confluence of the lesion-specific brain ROI. For simplicity, the lesion-specific brain ROI will be referred to as "brain" in the text. The individual anonymized dose-volume histogram (DVH) values were exported from MIM and saved as comma-separated value (CSV) files. In-house Python scripts were used to extract brain dose and volume metrics. The initial and recurrent lesions were evaluated individually. Binary radionecrosis data following re-irradiation SRS for recurrent lesions and single-session SRS for non-recurrent lesions were included in the modeling.

## NTCP modeling with time effect:

The physical dose metrics were converted into $EQD2_{\alpha/\beta}$ using the LQ-L model reported in our previous work [20] as described below:

$$EQD2_{\alpha/\beta} = \frac{BED}{\left(1 + \frac{2}{\alpha/\beta}\right)} \qquad (1)$$

Where,

$$BED = \begin{cases} nd + \frac{nd^2}{\alpha/\beta}, & if\ d < d_T \\ nd_T + \frac{nd_T^2}{\alpha/\beta} + \frac{\gamma}{\alpha}n(d - d_T), & if\ d \geq d_T \end{cases} \qquad (2)$$

Such that

$$d_T = 2\frac{\alpha}{\beta} \qquad (3)$$

$$\frac{\gamma}{\alpha} = 1 + \frac{2d_T}{\alpha/\beta} \qquad (4)$$

Where, $\gamma$ is log cell kill per Gy, and $d_T$ is the transition dose into the linear portion of the survival curve at higher doses. In this study, an approximate $\gamma/\alpha$ = 5 and $\alpha/\beta$ =3 were used based on prior studies [19, 21].

In-house Python scripts were used to develop a logistic model to predict radionecrosis risk following SRS. This logistic model, which uses dose or volume and radionecrosis risk as inputs, was validated as appropriate for NTCP modeling in the recent HyTEC paper on BMs [19]. In this model, NTCP is calculated as

$$NTCP = \frac{exp\left(4\gamma_{50}\left(\frac{D_x}{D_{x50}}-1\right)\right)}{1+exp\left(4\gamma_{50}\left(\frac{D_x}{D_{x50}}-1\right)\right)} \quad (5)$$

Where $D_x$ is the maximum dose to xcc of volume, $D_{x50}$ is the dose to xcc volume corresponding to a 50% risk of radionecrosis, and $\gamma_{50}$ is the slope parameter. The binary radionecrosis data, presented as a two-dimensional array of brain dose metrics and NTCP, was input into an in-house code to generate quantile bins for mean dose and outcome, along with Agresti-Coull [22] 95% confidence intervals ($CI_{95}$) estimation. The best-fit model parameters ($EQD2_{50}$ and $\gamma_{50}$) were generated by maximizing the log-likelihood function. We used the CMA-ES global optimization method, which belongs to the class of evolutionary algorithms provided by PyCMA [23]. It should be noted that the optimizer's choice is arbitrary, and any other method could be used. The quality of fit was assessed using Chi-2 ($\chi^2$).

In the first step, we fitted model parameters for the nonrecurrent (M0) and recurrent (M1-retreat) lesions. For the model, M1-combo of the recurrent dataset, the time-dependent dose discount factor for prior radiation was calculated using the Gompertzian bi-exponential decay function [24]. The Gompertzian function first introduced as an actuarial curve for an English population in 1883 [25], has found widespread application in pharmacodynamics, encompassing studies on pharmacokinetics, dose-response relationships, time effects on toxicity, and bone remodeling [24, 26-28]. A distinguishing feature of the Gompertzian function is the presence of two rate constants within an expo-exponential equation [24]. The rate constants can be flexibly used to characterize rapid and gradual growth or decay processes that may otherwise require formidable complexities of tissue response dynamics. In our M1-combo NTCP model, we employed the modified Gompertzian function to estimate the time-dependent dose-discount factor, DDF(t). The DDF at any time "t" is calculated using the formula:

$$DDF(t) = \frac{G(t)}{G(t)_{max}} \quad (6)$$

Where $G(t) = e^{(-k1(t-t_0)e^{(-k2(t-t_0))})}$ (7)

The DDF(t) normalization process scales the response to a maximum value of 1, facilitating comparative analysis across different parameter settings or time intervals. Specifically, in the current model, $k_1$ dictates the initial growth rate, $k_2$ controls the deceleration rate of growth, and "$t_0$" represents the time at which an event such as radionecrosis is initiated. The Figure 1 shows an example of the interplay of the $k_1$ and the $k_2$ constants.

As such, for the recurrent lesions, the combined brain dose at second irradiation is estimated as

$$D_{2cc}^{Total} = D_{2cc}^{Retreat} + D_{2cc}^{Initial} * (DDF(t)) + D_{2cc}^{WBRT} * (DDF(t)) \quad (8)$$

The constants $k_1$, $k_2$, and $t_0$ were estimated using the CMA-ES optimization algorithm. The parameters $k_1$ and $k_2$ were randomly initialized to 0.01 with bounds from 0 to 200, while $t_0$ was initialized to 60 days with bounds from 0 to 1000 days. This closely mirrored the radionecrosis onset observed in our non-recurrence dataset, ranging from 1.2 to 120 months, with an average onset at 10 months. The optimizer estimated the best-fit values to predict the dose metrics that would follow the outcome of M0. By minimizing the mean squared error between observed (M1-retreat) and predicted outcomes (M0), the algorithm iteratively refines these parameters until an optimal solution is achieved. This approach integrates numerical optimization methods with the Gompertzian function, offering a robust framework for parameter estimation and predictive modeling in the studied context. The best fit Gompertzian parameters were used to calculate the combined dose using Eq. 8, followed by creating the M1-combo NTCP model.

## Results:

The patient, disease, and treatment characteristics are summarized in Table 1. The BM lesions in the current cohort varied from 0.01 to 28.3cc, with brain $D_{0.2cc}$ doses ranging from 17 to 39 Gy. For non-recurrent BM lesions, the mean brain $D_{0.2cc}$ dose was 17.1 Gy, while for recurrent lesions, the initial and retreatment mean $D_{0.2cc}$ doses were 18.8 Gy. The time interval between initial and repeat SRS ranged from 1.9 to 102.9 months, with an average of 20.4 months. Some lesions (44) received whole-brain radiotherapy between the initial and retreat SRS.

Figure 2 shows the non-recurrence (M0) model results for various dose metrics. The solid circles represent the quantile bins of measured outcomes with error bars representing the Agresti-Coull approximate 95% confidence intervals. The M0 model fits were statistically significant as measured by goodness of fit statistics $\chi^2$ of 0.069, 0.039, and 0.058, respectively, for $D_{0.1cc}$, $D_{0.2cc}$, and $D_{0.3cc}$, and a p-value of ~1.000 indicating an insignificant difference between model prediction and the observed NTCP values. Given the minimal differences between these dosimetric variables, further modeling was focused on $D_{0.2cc}$, which had the most favorable $\chi^2$ values. Figure 3 illustrates the M1-retreat model, which confirms the robustness of the modeling methodology. With a $\chi^2$ of 0.089 and a p-value of 1.000, this model shows a statistically insignificant difference between the model predictions and observed NTCP values.

For the recurrence M1-combo model, the Gompertz parameters $k_1$, $k_2$, and $t_0$ of DDF were estimated by fitting the $D_{0.2cc}^{Total}$ and the observed outcome of the recurrence dataset to the expected outcome using $EQD2_{50}$ of 103.3 and $\gamma_{50}$ of 1.182 Gy. The fitting process involved CMA-ES optimization, iteratively refining solutions based on fitness evaluations until convergence and mean squared error were minimized. It generated best-fit values of $k_1$ = 0.870, $k_2$ = 0.012, and $t_0$ = 48.6 days. Henceforth, with the estimated brain $D_{0.2cc}^{Total}$, the observed recurrence outcome is fitted using the same approach as M0 model. The hollow circles in Figure 3 represent the quantile bins of estimated outcomes with error bars representing the Agresti-Coull approximate 95% confidence intervals.

The goodness of fit statistics, $\chi^2$, and p-value for M1-retreat was 0.055 and 1.000. Compared to the M1-retreat, the M1-combo predicted a higher threshold $EQD2_{50}$ of 165.1 Gy vs 87.6 Gy. Notably, the brain $D_{0.2cc}$ in M1-combo is expected to be higher due to cumulative dosing from initial and repeat SRS. A lower $\gamma_{50}$ value in M1-combo (0.48 Gy) compared to M1-retreat (0.58 Gy) indicates a more gradual radionecrosis risk response to dose changes, potentially allowing for flexibility in dose adjustments if an appropriately discounted initial dose is considered.

All three predicted NTCP models from Figure 2 and Figure 3 are shown together in Figure 4. In Table 2, we evaluated the models' NTCP predictions for two different brain $D_{0.2cc}$ doses. For a $D_{0.2cc}$ of 19 Gy, the mean dose delivered to the recurrent BM lesions, the predicted NTCP from best to worst was 0.08 in M0, 0.21 in M1-combo, and 0.28 in M1-retreat. We assessed the curves' steepness within our dataset's mean and maximum brain $D_{0.2cc}$ dose range by comparing the predicted NTCP differences between 19 Gy and 29 Gy doses ($NTCP_{29Gy}$ - $NTCP_{19Gy}$). The most considerable NTCP difference was observed in M1-retreat (0.16), followed by M0 (0.14), and the smallest in M1-combo (0.06). Alternatively, when looking at the dose needed for an NTCP change of 10% ($NTCP_{0.30}$-$NTCP_{0.20}$), the brain $D_{0.2cc}$ dose difference was highest for M1-combo (16 Gy), followed by M1-retreat (7 Gy), followed by M0 (4.3 Gy).

## Discussion:

While stereotactic radiosurgery (SRS) is recognized for effectively managing limited recurrent BMs [29, 30], it presents cautious outcomes with a two to threefold increase in necrosis risk [12, 31], potentially impacting neurological function and quality of life. These outcomes underscore the importance of predictive models such as Emami [32], HyTEC [19], and PenTEC [33], which provide guidelines for normal tissue response in conventional, hypo-fractionated, and pediatric dose scenarios, respectively. However, there remains a significant gap in similar models tailored for retreatment settings.

Our previous study investigated the variations in TCP for recurrent and non-recurrent lesions [20]. Building upon this, our current research investigates the impact of re-irradiation on NTCP, with a specific focus on the risk of radionecrosis. Notably, across all lesions in our dataset, we found that the logistic model aligns well with HyTEC recommendations, offering strong predictive capabilities alongside ease of implementation [19].

We investigated how NTCP models differ for recurrent BMs requiring repeat SRS (M1-retreat) compared to non-recurrent lesions (M0). Conceivably, the normal tissue adjacent to recurrent lesions may have reduced radiation tolerance due to incomplete recovery from prior treatment. Lower $EQD2_{50}$ and $\gamma_{50}$ values in M1-retreat (87.6 Gy, 0.58) than in M0 (103.3 Gy, 1.18) suggest greater radiosensitivity and a flatter dose-response curve. This implies a gradual increase in complication probability with the dose for M1-retreat, necessitating conservative dose constraints but allowing more flexible treatment adjustments than for M0, which requires higher doses with precise delivery to avoid a rapid increase in the risk of radionecrosis. Additionally, in M1-combo, incorporating time-dependent discounted doses had a $\gamma_{50}$ of 0.48, indicating a more gradual tissue response. The relatively flatter response of the M1-combo model suggests that when incorporating the influence of the initial SRS1 dose over time, the tissue appears less sensitive to the cumulative dose than when considering the repeat SRS dose alone. This underscores the complexity of tissue sensitivity influenced by the temporal distribution of doses beyond immediate dose considerations in optimizing therapeutic outcomes.

This study has a few limitations. This study included 149 recurrent BM lesions, one of the largest reported numbers in the literature, following HYTEC reporting standards [19]. However, this sample size is insufficient to fully assess the impact of the time interval between initial and repeat SRS, and other potential factors such as histology, patient age, and systemic therapies were not accounted for. More extensive pooled databases or prospective clinical trials would provide better opportunities for such analyses.

Normal tissue tolerance for reirradiation depends on the initial dose, the interval between treatments, and tissue type. Rapidly proliferating tissues, such as skin, exhibit early damage and repair, while slowly responding tissues, like the spinal cord, manifest responses over months to years. Assessing reirradiation tolerance necessitates considering the timing and extent of proliferative regeneration and the residual damage post-regeneration [34]. Our modified Gompertzian function estimates the DDF parameters $k_1$ and $k_2$ to simulate damage and repair tissue dynamics, with $t_0$ indicating damage initiation. Such an approach can potentially be adapted to model normal tissue response for other cancer sites.

It should be noted that this approach assumes similar tissue sensitivity between recurrent and non-recurrent BM lesions when accounting for DDF(t). The extent of tissue plasticity, or the ability of tissues to adapt and recover, is not well understood. A critical area for further investigation is whether repair mechanisms for sub-lethal radiation damage are altered by reirradiation at different temporal phases of tissue damage.

Additionally, like any model estimating reality, the validity of our approach is limited to the range within which it has been verified, and caution should be exercised when extrapolating beyond this range.

## Conclusions:

Using the largest standardized recurrent and non-recurrent BM SRS datasets, we develop and validate NTCP models that incorporate dose discounts and time-dependent recovery effects, which are crucial for understanding radionecrosis risk. The modified Gompertzian function effectively captured these radionecrosis risk dynamics, offering a framework for refining reirradiation-induced toxicity prediction. Our findings highlight the importance of tailored dosing strategies to optimize therapeutic outcomes and minimize adverse effects in recurrent lesions. Larger prospective studies are needed to validate these models across diverse patient populations and account for other prognostic factors, enhancing their precision and applicability in clinical practice.

**Table 1:** Summary of non-recurrent and recurrent patient cohort characteristics. Additional details are provided in references [12, 13].

| Patients | Non-retreatment | | Retreatment-1st SRS | | Retreatment-2nd SRS | |
|---|---|---|---|---|---|---|
| | mean±std | range | mean±std | range | mean±std | range |
| Male/Female | 99/163 | - | 26/61 | - | 26/61 | - |
| Age at first SRS (yrs) | 56.9 ± 12.4 | 14.5-86.0 | 52.9 ± 12.7 | 23.6-85.1 | 54.5 ± 12.8 | 27.2-86.5 |
| Lesion volume (cc) | 1.2 ± 2.7 | 0.01-24.5 | 1.8 ± 3.7 | 0.02-28.3 | 1.7 ± 2.8 | 0.02-23.7 |
| Primaries | | | | | | |
| Lung | 394 | | 31 | | 31 | |
| Breast | 282 | | 67 | | 67 | |
| Melanoma | 220 | | 33 | | 33 | |
| Gastrointestinal | 19 | | 8 | | 8 | |
| Other | 133 | | 10 | | 10 | |
| Brain $D_{0.2cc}$ (Gy) | 17.1 ± 4.1 | 0.1-39.3 | 18.8 ± 3.6 | 7.9-32.7 | 18.8 ± 3.4 | 9.6-28.5 |
| Brain $V_{12Gy}$ (cc) | 1.6 ± 2.5 | 0.1-19.3 | 2.2 ± 3.2 | 0.1-16.5 | 2.5 ± 2.8 | 0.1-15.1 |
| Time interval between 1st SRS and 2nd SRS (months) | – | – | – | – | 20.4 ± 16.4 | 1.9-102.9 |

**Table 2:** The model parameters and predicted differences in the NTCP models

| Model | EQD2$_{50}$ (Gy) | $\gamma_{50}$ (Gy) | Brain D$_{0.2cc}$ (Gy) | | NTCP for Brain D$_{0.2cc}$ | |
|---|---|---|---|---|---|---|
| | | | NTCP$_{0.20}$ | NTCP$_{0.30}$ | 19 Gy | 29 Gy |
| M0 | 103.3 (101.4, 105.4) | 1.18 (0.95, 1.53) | 27.1 | 31.4 | 0.08 | 0.22 |
| M1-retreat | 87.6 (64.5, 107.8) | 0.58 (0.34, 0.90) | 14.1 | 21.1 | 0.28 | 0.44 |
| M1-combo | 165.1 (111.2, 206.5) | 0.48 (0.27, 0.79) | 17.7 | 33.7 | 0.21 | 0.27 |

## List of Figures

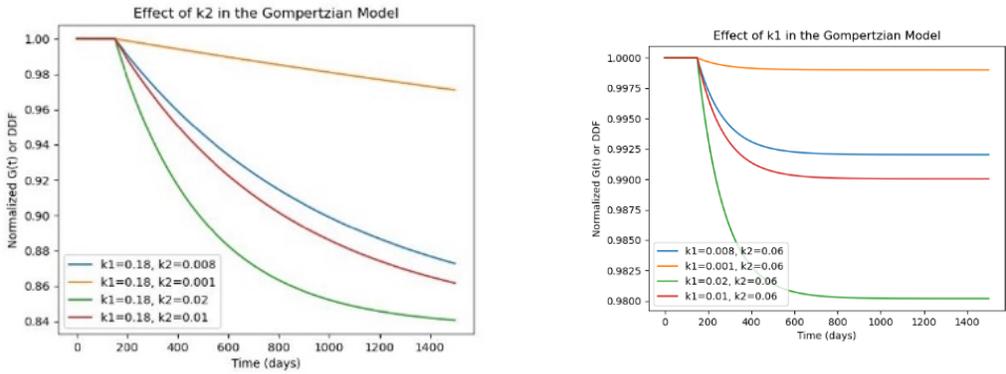

**Figure 1:** Simulation to show the effect of the two rate constants (k$_1$ and k2) in the DDF(t).

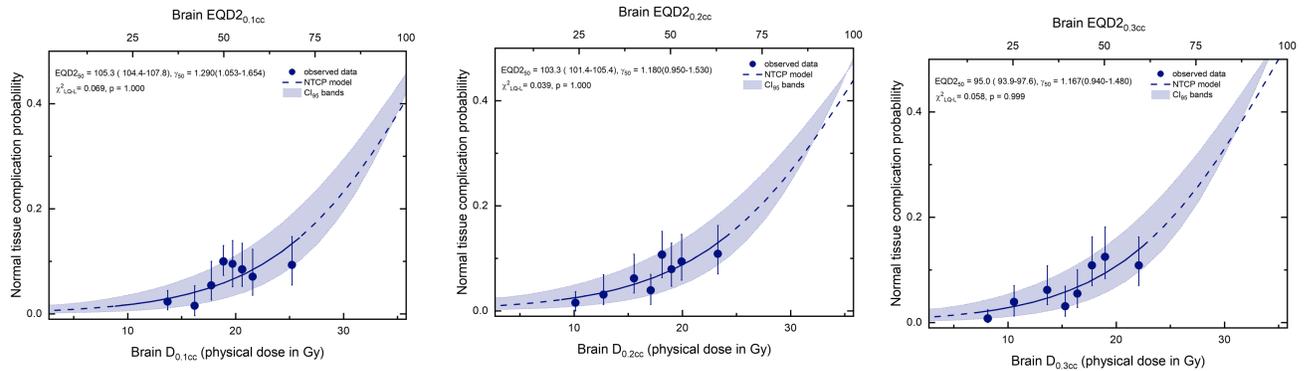

**Figure 2:** The one-year NTCP M0 model for non-recurrent BM lesion specific doses to 0.1cc (left), 0.2cc (middle) and 0.3cc (right) of brain tissue. The circle size is weighted by the lesion count and the error bars represent the weighted CI$_{95}$ upper and lower bounds. The dotted lines and bands represent the model fit for CI$_{95}$ bounds. The model fitted NTCP in the region outside data is shown with dotted dashed lines. The EQD2 dose conversion on the x-axis was done with the LQ-L model with an $\alpha/\beta$ of 3.

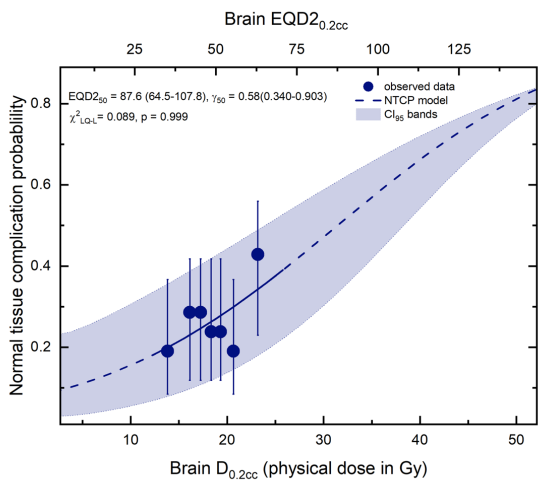 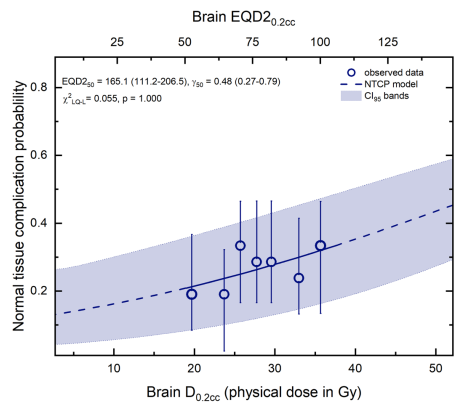

**Figure 3:** The one-year NTCP for recurrent brain metastases using M1-retreat and M1-combo models. The circle size is weighted by the lesion count and the error bars represent the weighted $CI_{95}$ upper and lower bounds. The dotted lines and bands represent the model fit for $CI_{95}$ bounds. The model fitted NTCP in the region outside data is shown with dotted dashed lines. The EQD2 dose conversion on the x-axis was done with the LQ-L model with an $\alpha/\beta$ of 3.

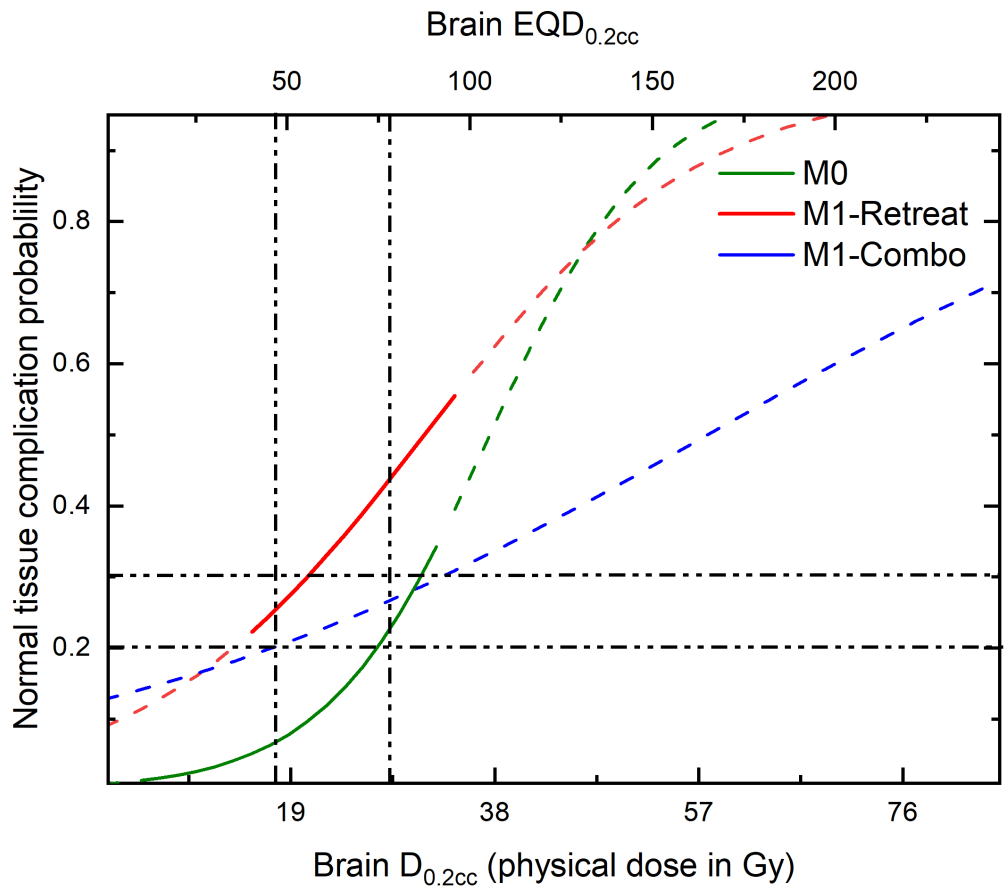

**Figure 4:** The predicted on-year NTCP model for single-fraction BM lesions in three different models. The horizontal dashed lines denote the model predicted brain $D_{2cc}$ for a NTCP of 0.20 and 0.30. The vertical lines denote the predicted NTCP for brain $D_{2cc}$ of 19 Gy and 29 Gy.